\documentclass[reprint, amsmath, amssymb, aps, prl, floatfix,superscriptaddress]{revtex4-2}

\usepackage{graphicx}
\usepackage{dcolumn}
\usepackage{bm}
\usepackage{booktabs}
\usepackage{multirow}
\usepackage{makecell}

\begin{document}
	
	\title{Weyl-Transition-Driven Giant Reversible Orbital Hall Conductivity}

	\author{Bo Zhao}
	\affiliation{Technical University of Darmstadt, 64287 Darmstadt, Germany}

	\author{Hao Wang}
	\email{haowang@tmm.tu-darmstadt.de}
	\affiliation{Technical University of Darmstadt, 64287 Darmstadt, Germany}

	\author{Wei Ren}
	\email{renwei@shu.edu.cn}
	\affiliation{Institute for Quantum Science and Technology, State Key Laboratory of Advanced Refractories, Materials Genome Institute, Physics Department, Shanghai University, Shanghai 200444, China}

	\author{Hongbin Zhang}
	\affiliation{Technical University of Darmstadt, 64287 Darmstadt, Germany}
	\date{\today}
	
	\begin{abstract}
		Orbital Hall conductivity (OHC) is a central ingredient of orbitronics, yet how to control it microscopically remains largely unexplored. Here we identify a general mechanism in which tilted Weyl crossings formed by orbitally distinct bands generate a strongly asymmetric orbital Berry curvature (OBC) distribution, whose imbalance survives Brillouin-zone integration and yields a sizable OHC already at zeroth order. Using first-principles calculations, we show that monolayer PtBi$_2$ realizes this mechanism and hosts a giant OHC dominated by a type-II Weyl point. A small biaxial tensile strain drives a type-II $\rightarrow$ type-I $\rightarrow$ type-II Weyl transition, leading to a reversible sign change of the OHC through the evolution of the OBC imbalance. This process is governed by the chiral orbital texture of the crossing bands and is further assisted by a strain-induced first-order structural phase transition through bonding reconstruction and polarization change. Our results establish Weyl engineering of orbital quantum geometry as a powerful route to generating and reversibly controlling OHC in polar multi-orbital materials.
	\end{abstract}
	\maketitle

	\begin{figure}[htb]
		\centering
		\includegraphics[width=0.9\linewidth]{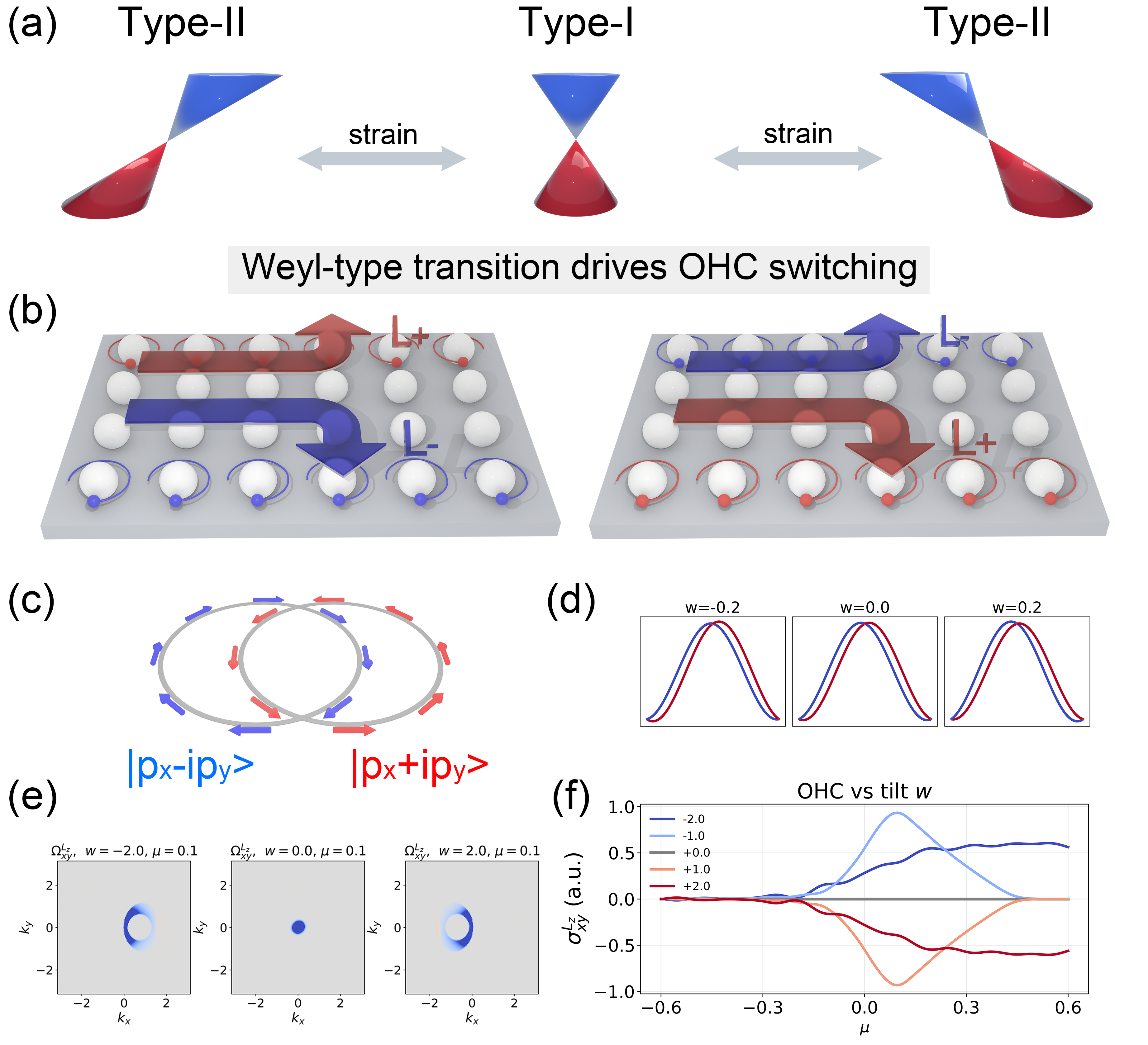}
		\caption{
			Weyl-point tilting–driven orbital Hall switching in the orbital Rashba model.
			(a) Schematic illustration of the strain-induced transition between different types of Weyl points.
			(b) Illustration of the OHC switching direction after the change in Weyl-point type.
			(c) Schematic orbital texture showing opposite chiral states: clockwise blue arrows and counterclockwise red arrows represent the $p_x - i p_y$ and $p_x + i p_y$ states, respectively.
			(d) Band structure of the orbital Rashba model with tilted Weyl points. Red and blue colors denote projections onto states with opposite chirality. The parameter $w$ controls the degree of Weyl-point tilting.
			(e) Distribution of OBC in the tight-binding (TB) model as a function of $w$.
			(f) Evolution of OHC in the orbital Rashba TB model as a function of Weyl-point tilting, where the color changes from blue to red correspond to $w$ varying from $-2$ to $2$.
		}
		\label{fig:fig1_schema}
	\end{figure}
	
	\noindent\textit{Introduction.—}Orbital angular momentum (OAM), alongside charge and spin, is a fundamental degree of freedom in atoms. In solids, however, OAM is generally believed to be largely quenched by the crystal field and to emerge only weakly through spin-orbit coupling (SOC)\cite{choi2023observation,go2018intrinsic}. Nevertheless, recent experiments have shown that orbital responses can exceed spin responses, giving rise to the emerging field of orbitronics, which is particularly promising for the electrical control of magnetism\cite{ding2022observation,choi2023observation,gao2024control,fert2024electrical,ding2024orbitala}. A central challenge in this field is to distinguish orbital torque generated by the orbital Hall effect (OHE) from spin torque induced by the spin Hall effect (SHE)\cite{go2020orbital,lee2021orbital,chen2025quantum,chen2024topologyengineered,ding2024orbitala,hayashi2023observation,sala2022giant,tanaka2008intrinsica,zheng2020magnetization}. This calls for identifying orbital Hall conductivity (OHC) in settings where it is linked to distinct physical properties beyond conventional spin transport, particularly through its interplay with ferroelectricity, magnetism, and band topology.

	In intrinsic SHE systems, large responses are often associated with SOC-gapped band crossings that generate pronounced spin Berry curvature (SBC). In contrast, giant OHC can arise even in light-element systems with negligible SOC\cite{jo2018gigantic,choi2023observation}, indicating that the orbital structure itself may play a more important role than SOC-induced band gaps. Although OHC and SHC share a common quantum geometric origin, the OHE is considerably richer because the OBC depends not only on the band Berry curvature (BC), but also on orbital hybridization, bonding configuration, and orbital texture\cite{lee2021orbital,go2018intrinsic,hagiwara2025orbital}. Therefore, identifying band crossings with distinct orbital character offers a general design principle for engineering enhanced OHC.

	This perspective becomes especially powerful in topological semimetals, where tilted Weyl crossings naturally generate highly structured momentum-space distributions of OBC\cite{soluyanov2015typeii}. Tilted Weyl crossings are known from the nonlinear Hall effect to host asymmetric BC distributions in momentum space when inversion symmetry is broken\cite{zhao2023berry,zhang2018electrically,kang2019nonlinear,du2021nonlinear,ma2019observation,sodemann2015quantum,zhang2021terahertz}, allowing a finite Berry-curvature dipole (BCD) even though the ordinary BC satisfies $\Omega(\mathbf{k})=-\Omega(-\mathbf{k})$ and vanishes upon Brillouin-zone integration. For the OHE, however, the consequence is more direct: the OBC itself can contribute to the OHC already at zeroth order. The OBC relevant to the OHE is generally even under time-reversal symmetry, $\Omega^{L}(\mathbf{k})=\Omega^{L}(-\mathbf{k})$, and thus need not vanish in the Brillouin zone. When redistributed by tilted Weyl crossings between bands of distinct orbital character, it can generate a sizable OHC already at the zeroth order. Tilted multi-orbital Weyl fermions therefore provide a natural platform for revealing and controlling the OHE.

	Guided by this mechanism, we seek materials that combine three key ingredients: complex orbital composition, tilted Weyl crossings, and broken inversion symmetry. Monolayer PtBi$_2$ emerges as an appealing candidate. This recently synthesized trigonal polar semimetal possesses strong orbital hybridization and a noncentrosymmetric structure, and has already attracted interest because of its superconducting surface Fermi arcs, giant shift current, and catalytic activity\cite{yang2024twodimensional,kvitnitskaya2025pointcontact,veyrat2025room,huang2025sizable,pan2026tunable,du2025atomically,hoffmann2025fermi,cai2025investigation,liu2025electrocatalytic,veyrat2023berezinskii,kuibarov2024evidence,changdar2025topological,gao2018possible,schimmel2024surface,vocaturo2024electronic,palumbo2025gapless,waje2025ginzburglandau}. These features suggest that PtBi$_2$ hosts an unusually rich electronic structure, making it particularly suitable for exploring the interplay among orbital quantum geometry, ferroelectricity, and Weyl physics.

	Here we show that monolayer PtBi$_2$ exhibits a giant OHC dominated by a type-II Weyl point. More remarkably, a small biaxial tensile strain drives a type-II $\rightarrow$ type-I $\rightarrow$ type-II Weyl transition and reverses the sign of the OHC, as schematically illustrated in Figs.~\ref{fig:fig1_schema}(a) and \ref{fig:fig1_schema}(b). We show that the Rashba-SOC-induced band opening preserves a type-II Weyl point that is highly sensitive to strain, while the Weyl transition strongly reshapes the orbital ordering and the momentum-space distribution of OBC.

	Furthermore, we identify a coupled first-order structural phase transition that facilitates this Weyl transition. Across the transition, the $z$-distance of the buckled Bi layer increases abruptly, accompanied by a sudden change in ferroelectric polarization. This structural and electronic reconstruction lifts the degeneracy, suppresses the electronic instability, and provides an additional driving force for the Weyl transition. Together, these results establish monolayer PtBi$_2$ as an attractive platform for strain-controlled orbitronics.
		 
	\begin{figure}[htb]
		\centering
		\includegraphics[width=0.9\linewidth]{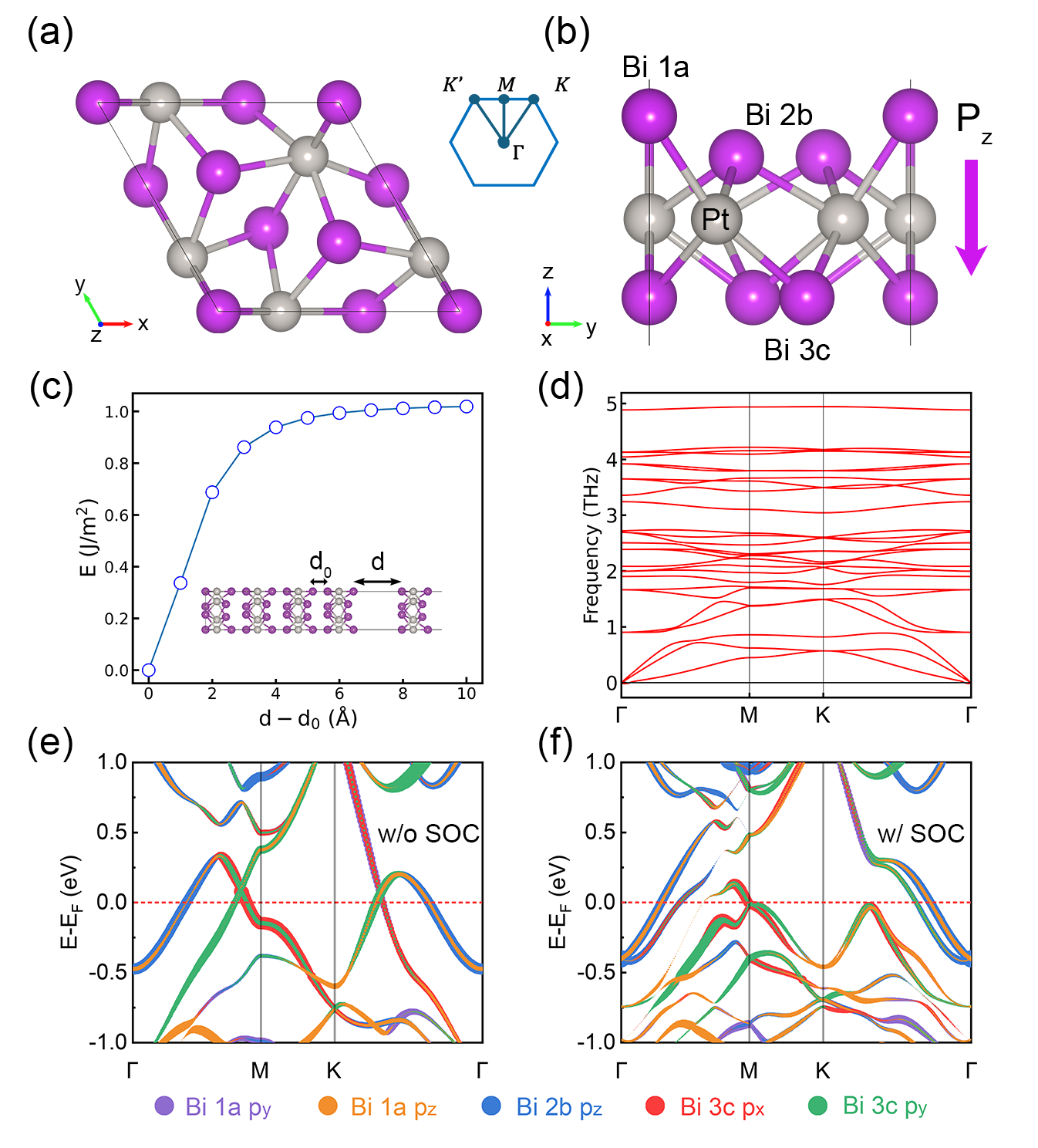}
		\caption{
			Crystal and electronic structure of monolayer PtBi$_2$.
			(a) Top view of the monolayer PtBi$_2$ crystal structure. The inset shows the Brillouin zone.
			(b) Side view of the monolayer structure. The central silver atoms correspond to Pt. The buckled structure consists of Bi atoms occupying Wyckoff 1a (top layer), 2b (second layer), and 3c (bottom layer) sites. The overall out-of-plane polarization points from the 1a layer toward the 3c layer.
			(c) Calculated exfoliation energy of monolayer PtBi$_2$. The inset illustrates the five-layer structure used in the calculation.
			(d) Phonon spectrum of monolayer PtBi$_2$.
			(e),(f) Orbital-resolved band structures of monolayer PtBi$_2$ without and with SOC, respectively. For clarity, only Bi orbital components are shown. Purple, orange, blue, red, and green dots represent Bi 1a $p_y$, 1a $p_z$, 2b $p_z$, 3c $p_x$, and 3c $p_y$ orbitals, respectively. The red dashed line indicates the Fermi level.
		}
		\label{fig:fig2_struct_band}
	\end{figure}
		
	\noindent\textit{Results.—}We employ a minimal Rashba model to elucidate the correlation among orbital character, Weyl-point tilting, and orbital Hall response. The Hamiltonian is written as
	
	\begin{equation}
		\begin{split}
			H(\mathbf{k}) =
			-\Big[\varepsilon(k_x,k_y)+w\sin k_x\Big]\mathbb{I} \\
			-\alpha\sin k_y\,\sigma_x
			+\alpha\sin k_x\,\sigma_y \\
			-\Big(M+v_z\sin k_z\Big)\sigma_z \, .
		\end{split}
	\end{equation}
	
	Here, $\varepsilon(k_x,k_y)=2 t\left(2-\cos k_x-\cos k_y\right)$ describes the nearest-neighbor hopping on a square lattice, while the $w\sin k_x$ term introduces the tilt of the band crossing. The term $\alpha(\sin k_x \sigma_y-\sin k_y \sigma_x)$ represents an orbital Rashba-type coupling, which is enabled by inversion-symmetry breaking and generates the orbital pseudospin texture. Here $\alpha$ characterizes the coupling strength. The last term, $M+v_z\sin k_z$, acts as an orbital Zeeman splitting in each 2D slice and gives rise to a Weyl-type dispersion in the full 3D model\cite{niu2019mixed,hanke2017mixed}.

	\begin{figure*}[htb]
		\centering
		\includegraphics[width=0.9\linewidth]{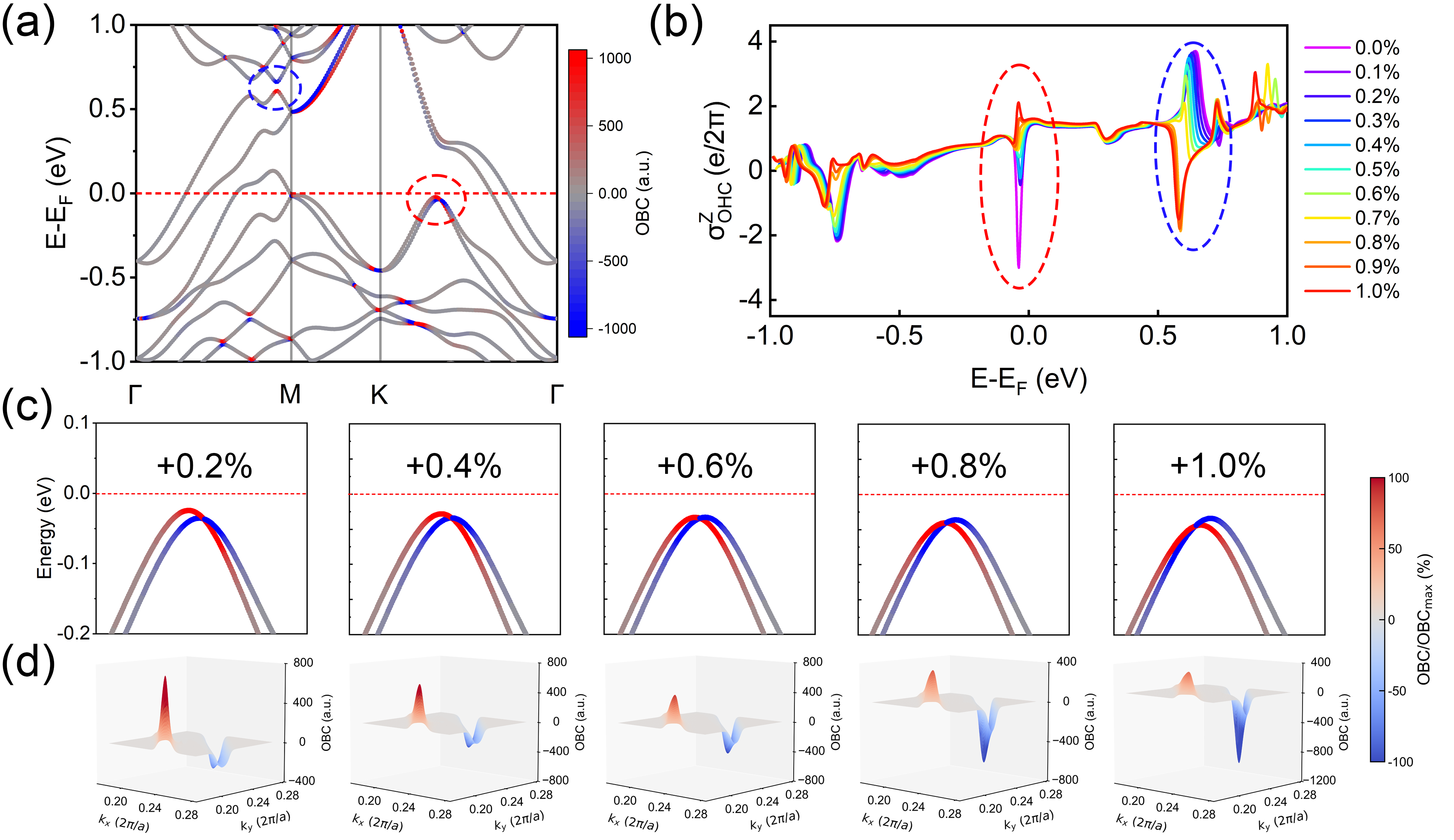}
		\caption{
				Strain-driven evolution of OBC and OHC.
				(a) $k$-resolved OBC projected onto the band structure. Red and blue colors represent positive and negative OBC, respectively. The dashed circles highlight the dominant OBC contributions near the Fermi level.
				(b) OHC as a function of biaxial tensile strain.
				(c) Strain-dependent band structure with $k$-resolved OBC along the K--$\Gamma$ path, focusing on the region highlighted by the red box.
				(d) Distribution of BC near the Weyl points corresponding to (c) under different strain values.
			}
		\label{fig:fig3_ohc_obc}
	\end{figure*}
	
	In the $(p_x,p_y)$ subspace, the OAM operator $L_z$ is represented by $\sigma_y$, whose eigenstates are the chiral orbital states $|p_x+i p_y\rangle$ and $|p_x-i p_y\rangle$ with opposite orbital angular momentum, as illustrated schematically in Fig.~\ref{fig:fig1_schema}(c). Figure~\ref{fig:fig1_schema}(d) shows the orbital-character-resolved band structure as a function of the tilt parameter $w$. The red and blue colors denote the projected orbital-polarization branches associated with $|p_x+i p_y\rangle$ and $|p_x-i p_y\rangle$, respectively. Changing the sign of $w$ shifts the tilted crossing from one side to the other and exchanges the order of the two orbital branches near the crossing. We emphasize that this evolution does not imply a reversal of the Weyl chirality or monopole charge. Instead, it only reflects a change in the projected chiral orbital character of the crossing bands. Since the crossing remains gapless throughout the evolution of $w$, this should not be regarded as a topological phase transition. This reversal strongly reshapes the momentum-space distribution of OBC, as shown in Fig.~\ref{fig:fig1_schema}(e), and consequently drives a sign change of the OHC in Fig.~\ref{fig:fig1_schema}(f).
	
	The above minimal model captures the key mechanism underlying the sign reversal of the OHC, namely the interplay among chiral orbital texture, Weyl-point tilting, and inversion-symmetry breaking. Motivated by this picture, we now turn to monolayer PtBi$_2$, where these ingredients naturally coexist.
	Monolayer PtBi$_2$ crystallizes in the $P31m$ space group, the same symmetry as its bulk counterpart. The optimized lattice constant is 6.517~\AA. 	Figure~\ref{fig:fig2_struct_band}(a) and~\ref{fig:fig2_struct_band}(b) show the top and side views of monolayer PtBi$_2$, respectively. Its structure consists of a distorted Bi--Pt--Bi sandwich, with the Pt atom located in the middle layer at the 3$c$ Wyckoff position. The six Bi atoms can be divided into three groups: the bottom three 3$c$ Bi atoms lie in the same plane, while one 1$a$ and two 2$b$ Bi atoms are located on the opposite side. The buckled 1$a$--2$b$ Bi layers break inversion symmetry, giving rise to an out-of-plane ferroelectric polarization.
		
 	This polarization points toward the 3$c$ side. As shown in Fig.~\ref{fig:fig2_struct_band}(c), the exfoliation energy is calculated to be 1.06~J/m$^2$, consistent with the previous report\cite{yang2024twodimensional}. As shown in Fig.~\ref{fig:fig2_struct_band}(d), the phonon spectrum contains no imaginary modes, confirming the dynamical stability of the ferroelectric state. By contrast, soft $B_{1u}$ and $B_{2u}$ phonon modes appear in the inversion-symmetric transition state.
		
	\begin{figure*}[htb]
		\centering
		\includegraphics[width=0.9\linewidth]{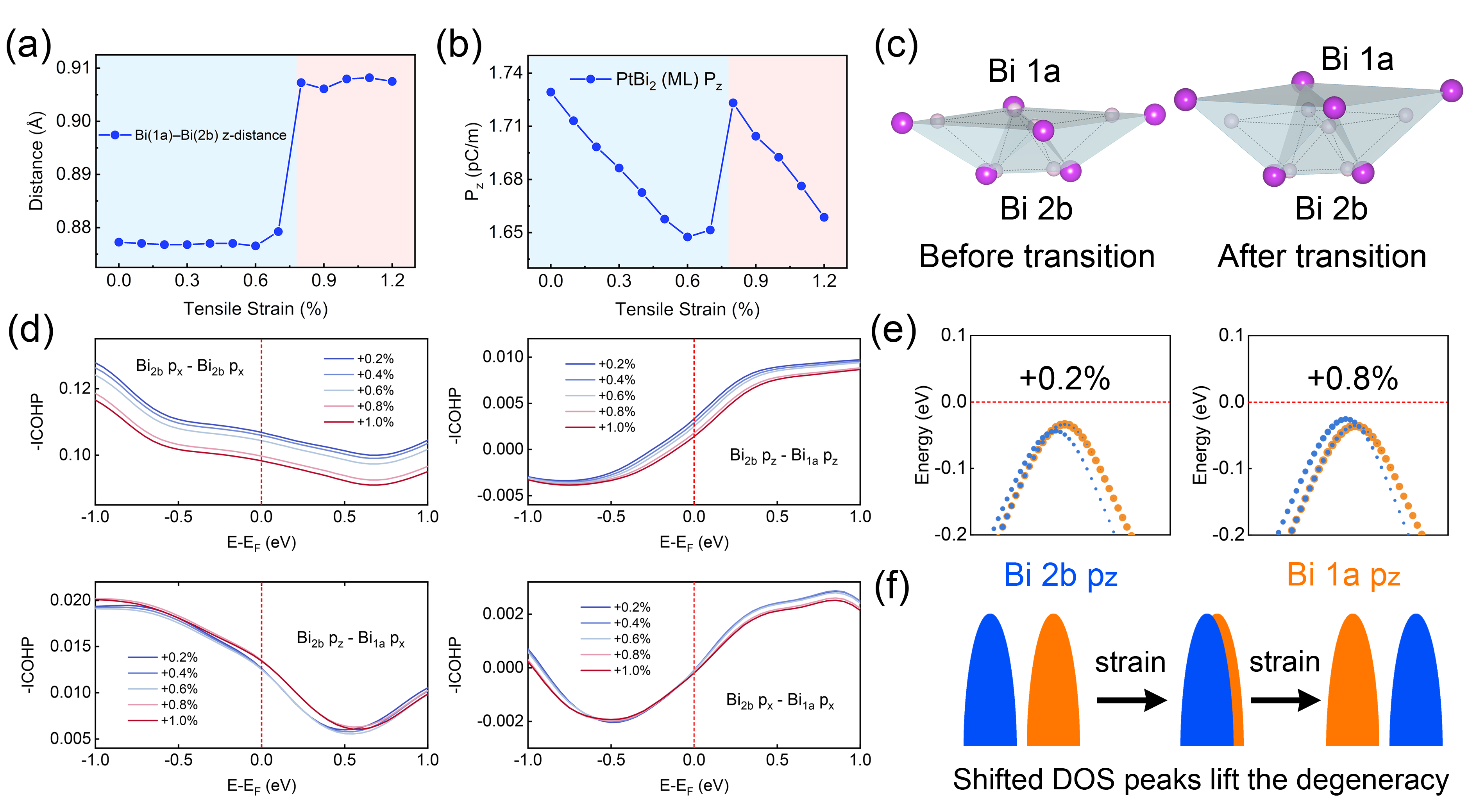}
		\caption{
			Structural phase transition and its electronic origin.
			(a) Strain dependence of the out-of-plane thickness between Bi 1a and Bi 2b atoms, showing an abrupt jump between 0.7\% and 0.8\%.
			(b) Evolution of the ferroelectric polarization under biaxial strain.
			(c) Schematic side view of the buckled Bi structure before and after the transition. The deformation is exaggerated for clarity; the dashed box marks the original configuration.
			(d) Strain-dependent $-$ICOHP for representative bonds: Bi 2b $p_x$--Bi 2b $p_x$, Bi 1a $p_z$--Bi 2b $p_z$, Bi 2b $p_z$--Bi 1a $p_x$, and Bi 2b $p_x$--Bi 1a $p_x$.
			(e) Orbital projections of Bi 1a (orange) and Bi 2b $p_z$ (blue) states. A degeneracy emerges near the type-I Weyl point at $+0.6\%$ strain due to overlapping DOS peaks.
			(f) Schematic illustration showing that splitting of DOS peaks lowers the total energy and drives the structural transition.
		}
		\label{fig:fig4_phase_trans_icohp}
	\end{figure*}

	The orbital-resolved band structures without and with SOC are shown in Fig.~\ref{fig:fig2_struct_band}(e) and \ref{fig:fig2_struct_band}(f), respectively. The states near the Fermi level are mainly composed of Pt 5$d$ and Bi 6$p$ orbitals. For clarity, only the key components of the Bi 6$p$ orbitals are highlighted. Without SOC, several bands cross the Fermi level, reflecting the metallic nature of the system. Along the K--$\Gamma$ path, a Weyl crossing is formed by the Bi 1$a$ $p_z$ orbital and the Bi 3$c$ $p_x$/$p_y$ orbitals. Consistent with our minimal model, this orbital combination can support a chiral orbital texture.

	With SOC included, the Weyl crossing along the K--$\Gamma$ path becomes gapped by about 250 meV. At the same time, the valence bands develop a Rashba-like splitting, with the originally degenerate bands shifting horizontally in momentum space. In addition, a new type-II Weyl point emerges, involving the Bi 1$a$ $p_z$ orbital and a mixture of the Bi 2$b$ $p_z$ and Bi 3$c$ $p_x$/$p_y$ orbitals. As shown below, this tilted Weyl point provides the dominant contribution to the OBC and the resulting OHC.

	Figure~\ref{fig:fig3_ohc_obc}(a) shows the $k$-resolved distribution of the OBC. Nonvanishing OBC mainly appears around the Weyl point along the K--$\Gamma$ path and near the M point. However, around the M point the positive and negative OBC contributions are nearly degenerate and largely cancel each other, yielding only a negligible contribution to the OHC.

	In contrast, the Weyl point along the K--$\Gamma$ path contributes significantly to the OHC near the Fermi level. As shown in Fig.~\ref{fig:fig3_ohc_obc}(b), the purple curve represents the OHC without strain, reaching about $-3~e/2\pi$. With increasing tensile strain, the OHC gradually decreases and changes sign when the strain exceeds +0.6\%, indicating that strain provides an efficient means to manipulate the OHC in monolayer PtBi$_2$.

	To further understand the origin of the OHC, we examine the strain evolution of the band structure and OBC distribution near the crossing point. As illustrated in Figs.~\ref{fig:fig3_ohc_obc}(c) and~\ref{fig:fig3_ohc_obc}(d), at a strain of +0.2\%, the positive OBC peak is significantly larger than the negative one. With increasing tensile strain, this imbalance gradually decreases. When the strain reaches +0.6\%, the Weyl point evolves into a type-I configuration, in which the positive and negative OBC contributions become nearly symmetric. As a result, their integrated contributions nearly cancel, leading to an OHC close to zero.

	When the strain exceeds +0.6\%, the Weyl point shifts to the left side of the crossing, and the bands hosting positive and negative OBC exchange their order, with the red branch moving above the blue one. This redistribution is also reflected in the OBC profile: the left peak is suppressed, while the right peak is enhanced. After Brillouin-zone integration, the sign of the OHC is consequently reversed.

	Furthermore, we observe another strain-induced OHC sign reversal about 0.5 eV above the Fermi level. The corresponding band structures are highlighted by the blue circles in Fig.~\ref{fig:fig3_ohc_obc}(a) and ~\ref{fig:fig3_ohc_obc}(b). In contrast to the mechanism near the Fermi level, this signal originates from a band-gap closing and reopening process rather than from the tilt-induced imbalance of the Weyl point. The gap reopening does not produce a quantized OHC plateau because other metallic bands coexist in the same energy window.
	
	The above results establish that the OHC sign reversal is electronically driven by the strain-induced evolution of the tilted Weyl point and the resulting redistribution of OBC. To further understand why this change becomes abrupt near the critical strain, we next examine the structural response of monolayer PtBi$_2$.
	
 	We identify a first-order structural transition that accompanies the abrupt change in the electronic structure and facilitates the OHC inversion. As shown in Fig.~\ref{fig:fig4_phase_trans_icohp}(a), the out-of-plane thickness of the Bi 1$a$ and Bi 2$b$ layers remains nearly unchanged below +0.7\%, but increases abruptly above this threshold. In contrast, the $z$ position of the flat Bi 3$c$ layer approaches the Pt layer almost linearly. As a result, the ferroelectric polarization along $z$ shows a sudden jump, increasing from 1.65~pC/m at +0.7\% to 1.73~pC/m at +0.8\%, and then gradually decreasing with further strain, as shown in Fig.~\ref{fig:fig4_phase_trans_icohp}(b). This anomalous response is reminiscent of a negative Poisson's ratio\cite{yu2017negative}. The top Bi layer forms a boat-shaped motif, with Bi 1$a$ at the top and Bi 2$b$ at the bottom. With increasing strain, the height of this motif remains nearly unchanged at first and then rises abruptly once the weak $p_z$--$p_z$ bonding can no longer support the structure. Figures~\ref{fig:fig4_phase_trans_icohp}(c) schematically show this structural phase transition.

	To further clarify this structural reconstruction, we plot the integrated crystal orbital Hamilton population (ICOHP) as a function of tensile strain for four representative bonds in Figs.~\ref{fig:fig4_phase_trans_icohp}(d). The top-left panel shows the strongest ``boat-bottom'' Bi 2$b$ $p_x$--Bi 2$b$ $p_x$ bond, whose strength gradually decreases with increasing tensile strain. The top-right panel displays the Bi 1$a$ $p_z$--Bi 2$b$ $p_z$ bond, which is directly related to the layer thickness. Its $-$ICOHP remains small throughout and gradually approaches zero with increasing tensile strain. As a result, the vertical bonding support along the $z$ direction is further weakened, making it increasingly difficult to maintain the height of boat-shaped structure. The bottom-left panel reveals a strain-induced tilted Bi 2$b$ $p_z$--Bi 1$a$ $p_x$ bond that compensates for the weakened vertical bonding. This is reflected in the sudden increase of $-$ICOHP once the tensile strain exceeds the critical value. Finally, the bottom-right panel shows the interlayer Bi 2$b$ $p_x$--Bi 1$a$ $p_x$ bond, which is only weakly affected by strain. These bonding analyses point to a clear picture: increasing tensile strain weakens the vertical bonds, while electronic reorganization promotes the formation of tilted bonds that stabilize the structure. This structural reconstruction provides the lattice support needed for the Weyl-point-driven OHC inversion. Without this phase transition, tensile strain would drive one Rashba-split branch to overlap completely with the other, thereby preventing the type-II $\rightarrow$ type-I $\rightarrow$ type-II Weyl transition.
		
	Furthermore, we suggest that the enhanced density of states (DOS) associated with the type-I Weyl point may provide an additional driving force for the structural phase transition. As illustrated in Figs.~\ref{fig:fig4_phase_trans_icohp}(e) and ~\ref{fig:fig4_phase_trans_icohp}(f), the orange and blue dots denote the Bi 1$a$ $p_z$ and Bi 2$b$ $p_z$ orbitals, respectively. Before the transition, the Bi 1$a$ $p_z$ state lies above the Bi 2$b$ $p_z$ state, whereas their order is reversed after the transition. Near +0.6\% tensile strain, the two Rashba branches approach each other and form a type-I Weyl point. At the same time, the Rashba band extrema enhance the DOS, bringing the two orbital components close to degeneracy and thereby promoting an electronic instability. This behavior is reminiscent of a Jahn--Teller-like mechanism, in which near-degeneracy is lifted to lower the total energy. Unlike a conventional Jahn--Teller distortion, however, the boat-shaped structure does not undergo a symmetry change during this first-order structural phase transition.	

	\noindent\textit{Conclusion.—} We demonstrate a giant OHC in monolayer trigonal Weyl semimetal PtBi$_2$. Remarkably, a small biaxial tensile strain reverses the sign of the OHC through a strain-driven Weyl transition that strongly reshapes the OBC distribution near the Fermi level. We further show that Weyl crossings formed by orbitally distinct bands generate a strongly asymmetric OBC distribution, with the associated chiral orbital texture playing a central role in the OHC response. For tilted Weyl points, this OBC imbalance can survive Brillouin-zone integration and produce a sizable OHC already at the zeroth order. In addition, we identify a possible first-order structural phase transition under biaxial strain, accompanied by a sudden change in the out-of-plane ferroelectric polarization. This structural reconstruction arises from bonding rearrangement and provides an additional driving force for the Weyl transition. Our results establish monolayer PtBi$_2$ as a promising platform for exploring the interplay among orbital quantum geometry, Weyl topology, and ferroelectric polarization, and more broadly highlight how multi-orbital complexity and polar symmetry can be harnessed to control orbital Hall responses in topological materials.

	\noindent\textit{Calculation methods.—}First-principles calculations were performed using the all-electron full-potential linearized augmented plane-wave method implemented in the \textsc{FLEUR} code, within the Perdew--Burke--Ernzerhof generalized-gradient approximation\cite{perdew1996generalized}. SOC was included self-consistently unless otherwise specified. Maximally localized Wannier functions were constructed using \textsc{Wannier90}\cite{marzari1997maximally,marzari2012maximally}. The OHC was evaluated from the Wannier-interpolated Hamiltonian using the Kubo formula. Chemical bonding was analyzed using \textsc{LOBSTER} based on separate \textsc{VASP} calculations, from which the ICOHP was extracted.

	The OHC is defined as
	\begin{equation}
		\sigma_{ij}^{z}
		=
		-e \int_{\mathrm{BZ}} \frac{d^{2}\mathbf{k}}{(2\pi)^{2}}
		\sum_{n}
		f_{n}(\mathbf{k})\,
		\Omega_{n,ij}^{z}(\mathbf{k}),
	\end{equation}
	where \(\Omega_{n,ij}^{z}(\mathbf{k})\) is the OBC
	\begin{equation}
		\Omega_{n,ij}^{z}(\mathbf{k})
		=
		2\hbar^{2}\,\mathrm{Im}
		\sum_{m\neq n}
		\frac{
			\langle u_{n\mathbf{k}}|\hat{J}_{i}^{z}|u_{m\mathbf{k}}\rangle
			\langle u_{m\mathbf{k}}|\hat{v}_{j}|u_{n\mathbf{k}}\rangle
		}{
			\left(E_{n\mathbf{k}}-E_{m\mathbf{k}}+i\eta\right)^{2}
		},
	\end{equation}
	with \(|u_{n\mathbf{k}}\rangle\) the cell-periodic Bloch state of energy \(E_{n\mathbf{k}}\), \(\hat{v}_j\) the velocity operator, and
	\begin{equation}
		\hat{J}_{i}^{z}=\frac{1}{2}\left\{\hat{v}_{i},\hat{L}_{z}\right\}
	\end{equation}
	the orbital-current operator.

	\noindent\textit{Acknowledgments.—}The authors thank the computing time provided to them on the high-performance computer Lichtenberg at the NHR Centers NHR4CES at TU Darmstadt. This work is funded by the Deutsche Forschungsgemeinschaft (DFG, German Research Foundation) -- CRC 1487, ``Iron, upgraded!'' -- with project number 443703006. H.~Wang and H.~Zhang also acknowledge support from the Deutsche Forschungsgemeinschaft (DFG, German Research Foundation) under Project-ID 463184206 -- SFB 1548. W.~Ren acknowledges support from the National Natural Science Foundation of China (Grant Nos.~52130204, 12311530675), Shanghai Engineering Research Center for Integrated Circuits and Advanced Display Materials, High-Performance Computing Center, Shanghai Technical Service Center of Science and Engineering Computing, Shanghai University.
	\bibliographystyle{apsrev4-2}
	\bibliography{refs.bib}

@article{cai2025investigation,
  title = {Investigation on High-Order Planar {{Hall}} Effect in Trigonal {PtBi$_2$}},
  author = {Cai, Fangqi and Chi, Mingxi and Hu, Yingjie and Liu, Heyao and Chen, Yangyang and Jing, Chao and Ren, Wei and Wang, He},
  date = {2025-06-09},
  year = {2025},
  journal = {Applied Physics Letters},
  shortjournal = {Appl. Phys. Lett.},
  volume = {126},
  number = {23},
  pages = {233101},
  doi = {10.1063/5.0258340}
}

@article{changdar2025topological,
  title = {Topological Nodal I-Wave Superconductivity in {PtBi$_2$}},
  author = {Changdar, Susmita and Suvorov, Oleksandr and Kuibarov, Andrii and Thirupathaiah, Setti and Shipunov, Grigory and Aswartham, Saicharan and Wurmehl, Sabine and Kovalchuk, Iryna and Koepernik, Klaus and Timm, Carsten and Büchner, Bernd and Fulga, Ion Cosma and Borisenko, Sergey and van den Brink, Jeroen},
  date = {2025-11},
  year = {2025},
  journal = {Nature},
  shortjournal = {Nature},
  volume = {647},
  number = {8090},
  pages = {613--618},
  doi = {10.1038/s41586-025-09712-6}
}

@article{chen2024topologyengineered,
  title = {Topology-{{Engineered Orbital Hall Effect}} in {{Two-Dimensional Ferromagnets}}},
  author = {Chen, Zhiqi and Li, Runhan and Bai, Yingxi and Mao, Ning and Zeer, Mahmoud and Go, Dongwook and Dai, Ying and Huang, Baibiao and Mokrousov, Yuriy and Niu, Chengwang},
  date = {2024-04-24},
  year = {2024},
  journal = {Nano Letters},
  shortjournal = {Nano Lett.},
  volume = {24},
  number = {16},
  pages = {4826--4833},
  doi = {10.1021/acs.nanolett.3c05129}
}

@article{chen2025quantum,
  title = {Quantum {{Orbital-to-Spin Conversion}} by {{Ferroelectric Topological Switch}}},
  author = {Chen, Zhiqi and Bai, Yingxi and Zeer, Mahmoud and Go, Dongwook and Dai, Ying and Huang, Baibiao and Mokrousov, Yuriy and Niu, Chengwang},
  date = {2025-11-18},
  year = {2025},
  journal = {Nano Letters},
  shortjournal = {Nano Lett.},
  volume = {25},
  number = {48},
  pages = {16817--16823},
  doi = {10.1021/acs.nanolett.5c04152}
}

@article{choi2023observation,
  title = {Observation of the Orbital {{Hall}} Effect in a Light Metal {{Ti}}},
  author = {Choi, Young-Gwan and Jo, Daegeun and Ko, Kyung-Hun and Go, Dongwook and Kim, Kyung-Han and Park, Hee Gyum and Kim, Changyoung and Min, Byoung-Chul and Choi, Gyung-Min and Lee, Hyun-Woo},
  date = {2023-07},
  year = {2023},
  journal = {Nature},
  shortjournal = {Nature},
  volume = {619},
  number = {7968},
  pages = {52--56},
  doi = {10.1038/s41586-023-06101-9}
}

@article{ding2022observation,
  title = {Observation of the {{Orbital Rashba-Edelstein Magnetoresistance}}},
  author = {Ding, Shilei and Liang, Zhongyu and Go, Dongwook and Yun, Chao and Xue, Mingzhu and Liu, Zhou and Becker, Sven and Yang, Wenyun and Du, Honglin and Wang, Changsheng and Yang, Yingchang and Jakob, Gerhard and Kläui, Mathias and Mokrousov, Yuriy and Yang, Jinbo},
  date = {2022-02-10},
  year = {2022},
  journal = {Physical Review Letters},
  shortjournal = {Phys. Rev. Lett.},
  volume = {128},
  number = {6},
  pages = {067201},
  doi = {10.1103/PhysRevLett.128.067201}
}

@article{ding2024orbitala,
  title = {Orbital {{Torque}} in {{Rare-Earth Transition-Metal Ferrimagnets}}},
  author = {Ding, Shilei and Kang, Min-Gu and Legrand, William and Gambardella, Pietro},
  date = {2024-06-05},
  year = {2024},
  journal = {Physical Review Letters},
  shortjournal = {Phys. Rev. Lett.},
  volume = {132},
  number = {23},
  pages = {236702},
  doi = {10.1103/PhysRevLett.132.236702}
}

@article{du2021nonlinear,
  title = {Nonlinear {{Hall}} Effects},
  author = {Du, Z. Z. and Lu, Hai-Zhou and Xie, X. C.},
  date = {2021-08-26},
  year = {2021},
  journal = {Nature Reviews Physics},
  shortjournal = {Nat. Rev. Phys.},
  volume = {3},
  number = {11},
  pages = {744--752},
  doi = {10.1038/s42254-021-00359-6}
}

@article{du2025atomically,
  title = {Atomically {{Ordered PtBi2 Intermetallic}} as {{Catalyst}} for {{Ultrahigh Efficiency}} and {{Durability}} in {{Methanol Electro-Oxidation}}},
  author = {Du, Wei and Guo, Wenjin and Zhu, Chengxin and Zhang, Wei and Li, Guangfang and Zhao, Huiping and Chen, Rong},
  date = {2025},
  year = {2025},
  journal = {Advanced Functional Materials},
  shortjournal = {Adv. Funct. Mater.},
  volume = {35},
  number = {31},
  pages = {2424532},
  doi = {10.1002/adfm.202424532}
}

@article{fert2024electrical,
  title = {Electrical Control of Magnetism by Electric Field and Current-Induced Torques},
  author = {Fert, Albert and Ramesh, Ramamoorthy and Garcia, Vincent and Casanova, Fèlix and Bibes, Manuel},
  date = {2024-03-13},
  year = {2024},
  journal = {Reviews of Modern Physics},
  shortjournal = {Rev. Mod. Phys.},
  volume = {96},
  number = {1},
  pages = {015005},
  doi = {10.1103/RevModPhys.96.015005}
}

@article{gao2018possible,
  title = {A Possible Candidate for Triply Degenerate Point Fermions in Trigonal Layered {PtBi$_2$}},
  author = {Gao, Wenshuai and Zhu, Xiangde and Zheng, Fawei and Wu, Min and Zhang, Jinglei and Xi, Chuanying and Zhang, Ping and Zhang, Yuheng and Hao, Ning and Ning, Wei and Tian, Mingliang},
  date = {2018-08-14},
  year = {2018},
  journal = {Nature Communications},
  shortjournal = {Nat. Commun.},
  volume = {9},
  number = {1},
  pages = {3249},
  doi = {10.1038/s41467-018-05730-3}
}

@article{gao2024control,
  title = {Control of Dynamic Orbital Response in Ferromagnets via Crystal Symmetry},
  author = {Gao, Tenghua and Rüßmann, Philipp and Wang, Qianwen and Fukunaga, Riko and Hayashi, Hiroki and Go, Dongwook and Harumoto, Takashi and Tu, Rong and Zhang, Song and Zhang, Lianmeng and Mokrousov, Yuriy and Shi, Ji and Ando, Kazuya},
  date = {2024-12},
  year = {2024},
  journal = {Nature Physics},
  shortjournal = {Nat. Phys.},
  volume = {20},
  number = {12},
  pages = {1896--1903},
  doi = {10.1038/s41567-024-02648-0}
}

@article{go2018intrinsic,
  title = {Intrinsic {{Spin}} and {{Orbital Hall Effects}} from {{Orbital Texture}}},
  author = {Go, Dongwook and Jo, Daegeun and Kim, Changyoung and Lee, Hyun-Woo},
  date = {2018},
  year = {2018},
  journal = {Physical Review Letters},
  shortjournal = {Phys. Rev. Lett.},
  volume = {121},
  number = {8},
  pages = {086602},
  doi = {10.1103/PhysRevLett.121.086602}
}

@article{go2020orbital,
  title = {Orbital Torque: {{Torque}} Generation by Orbital Current Injection},
  author = {Go, Dongwook and Lee, Hyun-Woo},
  date = {2020},
  year = {2020},
  journal = {Physical Review Research},
  shortjournal = {Phys. Rev. Res.},
  volume = {2},
  number = {1},
  pages = {013177},
  doi = {10.1103/PhysRevResearch.2.013177}
}

@article{hagiwara2025orbital,
  title = {Orbital {{Topology}} of {{Chiral Crystals}} for {{Orbitronics}}},
  author = {Hagiwara, Kenta and Chen, Ying-Jiun and Go, Dongwook and Tan, Xin Liang and Grytsiuk, Sergii and Yang, Kui-Hon Ou and Shu, Guo-Jiun and Chien, Jing and Shen, Yi-Hsin and Huang, Xiang-Lin and Cojocariu, Iulia and Feyer, Vitaliy and Lin, Minn-Tsong and Blügel, Stefan and Schneider, Claus Michael and Mokrousov, Yuriy and Tusche, Christian},
  date = {2025},
  year = {2025},
  journal = {Advanced Materials},
  shortjournal = {Adv. Mater.},
  volume = {37},
  number = {27},
  pages = {2418040},
  doi = {10.1002/adma.202418040}
}

@article{hanke2017mixed,
  title = {Mixed {{Weyl}} Semimetals and Low-Dissipation Magnetization Control in Insulators by Spin-Orbit Torques},
  author = {Hanke, Jan-Philipp and Freimuth, Frank and Niu, Chengwang and Blügel, Stefan and Mokrousov, Yuriy},
  date = {2017},
  year = {2017},
  journal = {Nature Communications},
  shortjournal = {Nat. Commun.},
  volume = {8},
  number = {1},
  pages = {1479},
  doi = {10.1038/s41467-017-01138-7}
}

@article{hayashi2023observation,
  title = {Observation of Long-Range Orbital Transport and Giant Orbital Torque},
  author = {Hayashi, Hiroki and Jo, Daegeun and Go, Dongwook and Gao, Tenghua and Haku, Satoshi and Mokrousov, Yuriy and Lee, Hyun-Woo and Ando, Kazuya},
  date = {2023-02-06},
  year = {2023},
  journal = {Communications Physics},
  shortjournal = {Commun. Phys.},
  volume = {6},
  number = {1},
  pages = {32},
  doi = {10.1038/s42005-023-01139-7}
}

@article{hoffmann2025fermi,
  title = {Fermi {{Arcs Dominating}} the {{Electronic Surface Properties}} of Trigonal {PtBi$_2$}},
  author = {Hoffmann, Sven and Schimmel, Sebastian and Vocaturo, Riccardo and Puig, Joaquin and Shipunov, Grigory and Janson, Oleg and Aswartham, Saicharan and Baumann, Danny and Büchner, Bernd and van den Brink, Jeroen and Fasano, Yanina and Facio, Jorge I. and Hess, Christian},
  date = {2025},
  year = {2025},
  journal = {Advanced Physics Research},
  shortjournal = {Adv. Phys. Res.},
  volume = {4},
  number = {5},
  pages = {2400150},
  doi = {10.1002/apxr.202400150}
}

@online{huang2025sizable,
  title = {Sizable Superconducting Gap and Anisotropic Chiral Topological Superconductivity in the {{Weyl}} Semimetal {PtBi$_2$}},
  author = {Huang, Xiaochun and Zhao, Lingxiao and Schimmel, Sebastian and Besproswanny, Julia and Härtl, Patrick and Hess, Christian and Büchner, Bernd and Bode, Matthias},
  date = {2025-07-18},
  year = {2025},
  number = {arXiv:2507.13843},
  eprint = {2507.13843},
  url = {https://arxiv.org/abs/2507.13843},
  eprinttype = {arXiv},
  doi = {10.48550/arXiv.2507.13843},
  pubstate = {prepublished}
}

@article{jo2018gigantic,
  title = {Gigantic Intrinsic Orbital {{Hall}} Effects in Weakly Spin-Orbit Coupled Metals},
  author = {Jo, Daegeun and Go, Dongwook and Lee, Hyun-Woo},
  date = {2018-12-04},
  year = {2018},
  journal = {Physical Review B},
  shortjournal = {Phys. Rev. B},
  volume = {98},
  number = {21},
  pages = {214405},
  doi = {10.1103/PhysRevB.98.214405}
}

@article{kang2019nonlinear,
  title = {Nonlinear Anomalous {{Hall}} Effect in Few-Layer {WTe$_2$}},
  author = {Kang, Kaifei and Li, Tingxin and Sohn, Egon and Shan, Jie and Mak, Kin Fai},
  date = {2019},
  year = {2019},
  journal = {Nature Materials},
  shortjournal = {Nat. Mater.},
  volume = {18},
  number = {4},
  pages = {324--328},
  doi = {10.1038/s41563-019-0294-7}
}

@article{kuibarov2024evidence,
  title = {Evidence of Superconducting {{Fermi}} Arcs},
  author = {Kuibarov, Andrii and Suvorov, Oleksandr and Vocaturo, Riccardo and Fedorov, Alexander and Lou, Rui and Merkwitz, Luise and Voroshnin, Vladimir and Facio, Jorge I. and Koepernik, Klaus and Yaresko, Alexander and Shipunov, Grigory and Aswartham, Saicharan and van den Brink, Jeroen and Büchner, Bernd and Borisenko, Sergey},
  date = {2024-02},
  year = {2024},
  journal = {Nature},
  shortjournal = {Nature},
  volume = {626},
  number = {7998},
  pages = {294--299},
  doi = {10.1038/s41586-023-06977-7}
}

@online{kvitnitskaya2025pointcontact,
  title = {Point-Contact Enhanced Superconductivity in Trigonal {PtBi$_2$}: Quest for the Origin of High-{{Tc}}},
  author = {Kvitnitskaya, O. E. and Harnagea, L. and Shipunov, G. and Aswartham, S. and Fisun, V. V. and Efremov, D. V. and Büchner, B. and Naidyuk, Yu G.},
  date = {2025-11-02},
  year = {2025},
  number = {arXiv:2511.00920},
  eprint = {2511.00920},
  url = {https://arxiv.org/abs/2511.00920},
  eprinttype = {arXiv},
  doi = {10.48550/arXiv.2511.00920},
  pubstate = {prepublished}
}

@article{lee2021orbital,
  title = {Orbital Torque in Magnetic Bilayers},
  author = {Lee, Dongjoon and Go, Dongwook and Park, Hyeon-Jong and Jeong, Wonmin and Ko, Hye-Won and Yun, Deokhyun and Jo, Daegeun and Lee, Soogil and Go, Gyungchoon and Oh, Jung Hyun and Kim, Kab-Jin and Park, Byong-Guk and Min, Byoung-Chul and Koo, Hyun Cheol and Lee, Hyun-Woo and Lee, OukJae and Lee, Kyung-Jin},
  date = {2021-11-18},
  year = {2021},
  journal = {Nature Communications},
  shortjournal = {Nat. Commun.},
  volume = {12},
  number = {1},
  pages = {6710},
  doi = {10.1038/s41467-021-26650-9}
}

@article{liu2025electrocatalytic,
  title = {Electrocatalytic Hydrogen Evolution and Doping Modification of {{Weyl}} Semimetal {PtBi$_2$} in Acidic Solutions},
  author = {Liu, Yinghui and Chen, Qiubo and An, Yukai and Qiu, Hailong and Ma, Shihui and Liu, Hongjun and Hu, Zhanggui and Wu, Yicheng},
  date = {2025},
  year = {2025},
  journal = {Inorganic Chemistry Frontiers},
  shortjournal = {Inorg. Chem. Front.},
  volume = {12},
  number = {8},
  pages = {3217--3228},
  doi = {10.1039/D5QI00022J}
}

@article{ma2019observation,
  title = {Observation of the Nonlinear {{Hall}} Effect under Time-Reversal-Symmetric Conditions},
  author = {Ma, Qiong and Xu, Su-Yang and Shen, Huitao and MacNeill, David and Fatemi, Valla and Chang, Tay-Rong and Mier Valdivia, Andrés M. and Wu, Sanfeng and Du, Zongzheng and Hsu, Chuang-Han and Fang, Shiang and Gibson, Quinn D. and Watanabe, Kenji and Taniguchi, Takashi and Cava, Robert J. and Kaxiras, Efthimios and Lu, Hai-Zhou and Lin, Hsin and Fu, Liang and Gedik, Nuh and Jarillo-Herrero, Pablo},
  date = {2019},
  year = {2019},
  journal = {Nature},
  shortjournal = {Nature},
  volume = {565},
  number = {7739},
  pages = {337--342},
  doi = {10.1038/s41586-018-0807-6}
}

@article{marzari1997maximally,
  title = {Maximally Localized Generalized {{Wannier}} Functions for Composite Energy Bands},
  author = {Marzari, Nicola and Vanderbilt, David},
  date = {1997},
  year = {1997},
  journal = {Physical Review B},
  shortjournal = {Phys. Rev. B},
  volume = {56},
  number = {20},
  pages = {12847--12865},
  doi = {10.1103/PhysRevB.56.12847}
}

@article{marzari2012maximally,
  title = {Maximally Localized {{Wannier}} Functions: {{Theory}} and Applications},
  author = {Marzari, Nicola and Mostofi, Arash A. and Yates, Jonathan R. and Souza, Ivo and Vanderbilt, David},
  date = {2012},
  year = {2012},
  journal = {Reviews of Modern Physics},
  shortjournal = {Rev. Mod. Phys.},
  volume = {84},
  number = {4},
  pages = {1419--1475},
  doi = {10.1103/RevModPhys.84.1419}
}

@article{niu2019mixed,
  title = {Mixed Topological Semimetals Driven by Orbital Complexity in Two-Dimensional Ferromagnets},
  author = {Niu, Chengwang and Hanke, Jan-Philipp and Buhl, Patrick M. and Zhang, Hongbin and Plucinski, Lukasz and Wortmann, Daniel and Blügel, Stefan and Bihlmayer, Gustav and Mokrousov, Yuriy},
  date = {2019-07-18},
  year = {2019},
  journal = {Nature Communications},
  shortjournal = {Nat. Commun.},
  volume = {10},
  number = {1},
  pages = {3179},
  doi = {10.1038/s41467-019-10930-6}
}

@article{palumbo2025gapless,
  title = {Gapless Topological {{Peierls-like}} Instabilities in More than One Dimension},
  author = {Palumbo, Santiago and Cornaglia, Pablo S. and Facio, Jorge I.},
  date = {2025-11-20},
  year = {2025},
  journal = {Physical Review B},
  shortjournal = {Phys. Rev. B},
  volume = {112},
  number = {20},
  pages = {L201117},
  doi = {10.1103/x6zl-t5hs}
}

@online{pan2026tunable,
  title = {Tunable {{Edelstein}} Effect in Intrinsic Two-Dimensional Ferroelectric Metal {PtBi$_2$}},
  author = {Pan, Weiyi and Fabian, Jaroslav},
  date = {2026-01-23},
  year = {2026},
  number = {arXiv:2601.16980},
  eprint = {2601.16980},
  url = {https://arxiv.org/abs/2601.16980},
  eprinttype = {arXiv},
  doi = {10.48550/arXiv.2601.16980},
  pubstate = {prepublished}
}

@article{perdew1996generalized,
  title = {Generalized {{Gradient Approximation Made Simple}}},
  author = {Perdew, J. P. and Burke, K. and Ernzerhof, M.},
  date = {1996},
  year = {1996},
  journal = {Physical Review Letters},
  shortjournal = {Phys. Rev. Lett.},
  volume = {77},
  number = {18},
  pages = {3865--3868},
  doi = {10.1103/PhysRevLett.77.3865}
}

@article{sala2022giant,
  title = {Giant Orbital {{Hall}} Effect and Orbital-to-Spin Conversion in 3d, 5d, and 4f Metallic Heterostructures},
  author = {Sala, Giacomo and Gambardella, Pietro},
  date = {2022-07-13},
  year = {2022},
  journal = {Physical Review Research},
  shortjournal = {Phys. Rev. Res.},
  volume = {4},
  number = {3},
  pages = {033037},
  doi = {10.1103/PhysRevResearch.4.033037}
}

@article{schimmel2024surface,
  title = {Surface Superconductivity in the Topological {{Weyl}} Semimetal T-{PtBi$_2$}},
  author = {Schimmel, Sebastian and Fasano, Yanina and Hoffmann, Sven and Besproswanny, Julia and Corredor Bohorquez, Laura Teresa and Puig, Joaquín and Elshalem, Bat-Chen and Kalisky, Beena and Shipunov, Grigory and Baumann, Danny and Aswartham, Saicharan and Büchner, Bernd and Hess, Christian},
  date = {2024-11-15},
  year = {2024},
  journal = {Nature Communications},
  shortjournal = {Nat. Commun.},
  volume = {15},
  number = {1},
  pages = {9895},
  doi = {10.1038/s41467-024-54389-6}
}

@article{sodemann2015quantum,
  title = {Quantum {{Nonlinear Hall Effect Induced}} by {{Berry Curvature Dipole}} in {{Time-Reversal Invariant Materials}}},
  author = {Sodemann, Inti and Fu, Liang},
  date = {2015},
  year = {2015},
  journal = {Physical Review Letters},
  shortjournal = {Phys. Rev. Lett.},
  volume = {115},
  number = {21},
  pages = {216806},
  doi = {10.1103/PhysRevLett.115.216806}
}

@article{soluyanov2015typeii,
  title = {Type-{{II Weyl}} Semimetals},
  author = {Soluyanov, Alexey A. and Gresch, Dominik and Wang, Zhijun and Wu, QuanSheng and Troyer, Matthias and Dai, Xi and Bernevig, B. Andrei},
  date = {2015-11},
  year = {2015},
  journal = {Nature},
  shortjournal = {Nature},
  volume = {527},
  number = {7579},
  pages = {495--498},
  doi = {10.1038/nature15768}
}

@article{tanaka2008intrinsica,
  title = {Intrinsic Spin {{Hall}} Effect and Orbital {{Hall}} Effect in \$4d\$ and \$5d\$ Transition Metals},
  author = {Tanaka, T. and Kontani, H. and Naito, M. and Naito, T. and Hirashima, D. S. and Yamada, K. and Inoue, J.},
  date = {2008-04-11},
  year = {2008},
  journal = {Physical Review B},
  shortjournal = {Phys. Rev. B},
  volume = {77},
  number = {16},
  pages = {165117},
  doi = {10.1103/PhysRevB.77.165117}
}

@article{veyrat2023berezinskii,
  title = {Berezinskii--{{Kosterlitz}}--{{Thouless Transition}} in the {{Type-I Weyl Semimetal}} {PtBi$_2$}},
  author = {Veyrat, Arthur and Labracherie, Valentin and Bashlakov, Dima L. and Caglieris, Federico and Facio, Jorge I. and Shipunov, Grigory and Charvin, Titouan and Acharya, Rohith and Naidyuk, Yurii and Giraud, Romain and van den Brink, Jeroen and Büchner, Bernd and Hess, Christian and Aswartham, Saicharan and Dufouleur, Joseph},
  date = {2023-02-22},
  year = {2023},
  journal = {Nano Letters},
  shortjournal = {Nano Lett.},
  volume = {23},
  number = {4},
  pages = {1229--1235},
  doi = {10.1021/acs.nanolett.2c04297}
}

@online{veyrat2025room,
  title = {Room Temperature {{Planar Hall}} Effect in Nanostructures of Trigonal-{PtBi$_2$}},
  author = {Veyrat, Arthur and Koepernik, Klaus and Veyrat, Louis and Shipunov, Grigory and Kovalchuk, Iryna and Aswartham, Saicharan and Qu, Jiang and Kumar, Ankit and Ceccardi, Michele and Caglieris, Federico and Rodríguez, Nicolás Pérez and Giraud, Romain and Büchner, Bernd and van den Brink, Jeroen and Ortix, Carmine and Dufouleur, Joseph},
  date = {2025-10-24},
  year = {2025},
  number = {arXiv:2410.12596},
  eprint = {2410.12596},
  url = {https://arxiv.org/abs/2410.12596},
  eprinttype = {arXiv},
  doi = {10.48550/arXiv.2410.12596},
  pubstate = {prepublished}
}

@article{vocaturo2024electronic,
  title = {Electronic Structure of the Surface-Superconducting {{Weyl}} Semimetal {PtBi$_2$}},
  author = {Vocaturo, Riccardo and Koepernik, Klaus and Facio, Jorge I. and Timm, Carsten and Fulga, Ion Cosma and Janson, Oleg and van den Brink, Jeroen},
  date = {2024-08-05},
  year = {2024},
  journal = {Physical Review B},
  shortjournal = {Phys. Rev. B},
  volume = {110},
  number = {5},
  pages = {054504},
  doi = {10.1103/PhysRevB.110.054504}
}

@article{waje2025ginzburglandau,
  title = {Ginzburg-{{Landau}} Theory for Unconventional Surface Superconductivity in {PtBi$_2$}},
  author = {Waje, Harald and Jakubczyk, Fabian and van den Brink, Jeroen and Timm, Carsten},
  date = {2025-10-30},
  year = {2025},
  journal = {Physical Review B},
  shortjournal = {Phys. Rev. B},
  volume = {112},
  number = {14},
  pages = {144519},
  doi = {10.1103/kkqg-ntcz}
}

@article{yang2024twodimensional,
  title = {Two-{{Dimensional Topological Ferroelectric Metal}} with {{Giant Shift Current}}},
  author = {Yang, Liu and Li, Lei and Yu, Zhi-Ming and Wu, Menghao and Yao, Yugui},
  date = {2024-10-28},
  year = {2024},
  journal = {Physical Review Letters},
  shortjournal = {Phys. Rev. Lett.},
  volume = {133},
  number = {18},
  pages = {186801},
  doi = {10.1103/PhysRevLett.133.186801}
}

@article{yu2017negative,
  title = {Negative {{Poisson}}’s Ratio in {{1T-type}} Crystalline Two-Dimensional Transition Metal Dichalcogenides},
  author = {Yu, Liping and Yan, Qimin and Ruzsinszky, Adrienn},
  date = {2017-05-25},
  year = {2017},
  journal = {Nature Communications},
  shortjournal = {Nat. Commun.},
  volume = {8},
  number = {1},
  pages = {15224},
  doi = {10.1038/ncomms15224}
}

@article{zhang2018electrically,
  title = {Electrically Tuneable Nonlinear Anomalous {{Hall}} Effect in Two-Dimensional Transition-Metal Dichalcogenides {WTe$_2$} and {MoTe$_2$}},
  author = {Zhang, Yang and van den Brink, Jeroen and Felser, Claudia and Yan, Binghai},
  date = {2018},
  year = {2018},
  journal = {2D Materials},
  shortjournal = {2D Mater.},
  volume = {5},
  number = {4},
  pages = {044001},
  doi = {10.1088/2053-1583/aad1ae}
}

@article{zhang2021terahertz,
  title = {Terahertz Detection Based on Nonlinear {{Hall}} Effect without Magnetic Field},
  author = {Zhang, Yang and Fu, Liang},
  date = {2021},
  year = {2021},
  journal = {Proceedings of the National Academy of Sciences of the United States of America},
  shortjournal = {PNAS},
  volume = {118},
  number = {21},
  pages = {e2100736118},
  doi = {10.1073/pnas.2100736118}
}

@article{zhao2023berry,
  title = {Berry Curvature Dipole and Nonlinear {{Hall}} Effect in Two-Dimensional {Nb$_{2n+1}$Si$_n$Te$_{4n+2}$}},
  author = {Zhao, Yiwei and Cao, Jin and Zhang, Zeying and Li, Si and Li, Yan and Ma, Fei and Yang, Shengyuan A.},
  date = {2023-05-11},
  year = {2023},
  journal = {Physical Review B},
  shortjournal = {Phys. Rev. B},
  volume = {107},
  number = {20},
  pages = {205124},
  doi = {10.1103/PhysRevB.107.205124}
}

@article{zheng2020magnetization,
  title = {Magnetization Switching Driven by Current-Induced Torque from Weakly Spin-Orbit Coupled {{Zr}}},
  author = {Zheng, Z. C. and Guo, Q. X. and Jo, D. and Go, D. and Wang, L. H. and Chen, H. C. and Yin, W. and Wang, X. M. and Yu, G. H. and He, W. and Lee, H.-W. and Teng, J. and Zhu, T.},
  date = {2020-02-05},
  year = {2020},
  journal = {Physical Review Research},
  shortjournal = {Phys. Rev. Res.},
  volume = {2},
  number = {1},
  pages = {013127},
  doi = {10.1103/PhysRevResearch.2.013127}
}
	
\end{document}